# "Thermal Spike" model applied to thin targets irradiated with swift heavy ion beams at few MeV/u

*Christelle* Stodel[1,*], *Marcel* Toulemonde[2], *Christoph* Fransen[3], *Bertrand* Jacquot[1], *Emmanuel* Clément[1], and *Christian* Dufour[4]

[1]Grand Accélérateur National d'Ions Lourds (GANIL), CNRS/CEA, Bvd Henri Becquerel, BP 55027, 14076 Caen CEDEX 5, France
[2]Centre de Recherche sur les Ions, les Matériaux et la Photonique (CIMAP), Bvd Henri Becquerel, BP 5133, 14070 Caen CEDEX 4, France
[3]Institut fuer Kernphysik, Universitaet zu Koeln, Zuelpicher Strasse 77, 50937 Koeln, Germany
[4]Centre de Recherche sur les Ions, les Matériaux et la Photonique (CIMAP), 6 Bvd du Maréchal Juin, 14050 Caen CEDEX 4, France

**Abstract.** High electronic excitations in radiation of metallic targets with swift heavy ion beams at the coulomb barrier play a dominant role in the damaging processes of some metals. The inelastic thermal spike model was developed to describe tracks in materials and is applied in this paper to some systems beams/targets employed recently in some nuclear physics experiments. Taking into account the experimental conditions and the approved electron-phonon coupling factors, the results of the calculation enable to interpret the observation of the fast deformation of some targets.

## 1 Introduction

In nuclear physics, structural evolution of nuclei when moving (far) away from magic numbers [1] remains intriguing in many aspects. It is not well theoretically reproduced so far by state-of-the-art models. High resolution spectroscopic data reveal this nuclear structure evolution as a function of spin, angular momentum, excitation energy and isospin. Their comparison with advanced nuclear model allow a microscopically description of these evolutions.

High resolution gamma-ray spectroscopy sheds light to the nuclear structure by measuring energy of excited states, their decay branching ratios and their lifetime. The Recoil Doppler Distance-Shift (RDDS) technique is an outstanding method to extract the lifetimes of the states of interest to deduce absolute transition strengths that are essential to test theoretical approaches. Here we discuss target problems in measurements using a so-called plunger device for half-life measurements using the RDDS technique [2] combined with the VAMOS++ spectrometer and the tracking germanium detector array AGATA (Fig. 1).

In these recent studies, exotic (i.e. neutron deficient or neutron rich) nuclei of interest were produced in inverse kinematics either by deep inelastic or by fusion-fission and fusion-evaporation reaction using heavy ion beams around the Coulomb barrier.

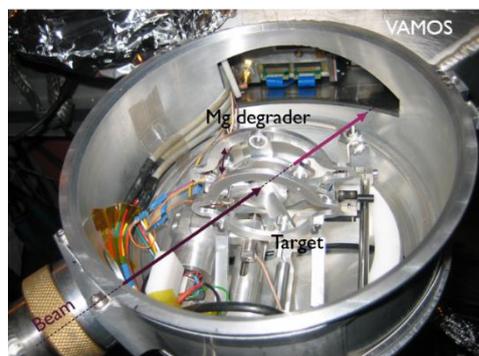

*Fig. 1. The Plunger set-up in the AGATA-VAMOS vacuum chamber.*

Contrary to previous experiments using a similar set-up but exploring other nuclei produced in direct kinematics, it was observed that some targets were slightly or seriously deformed even at very low heavy ion beam current (< 1 pnA). Unfortunately, this shape change is at the detriment of the quality of measurements: the distance between the plunger target and a plunger degrader foil used to slow down the recoils should be known and constant with a precision of better than 1 µm but the wrinkles height of the material was observed to be of the order of 100 µm, see Fig. 2.

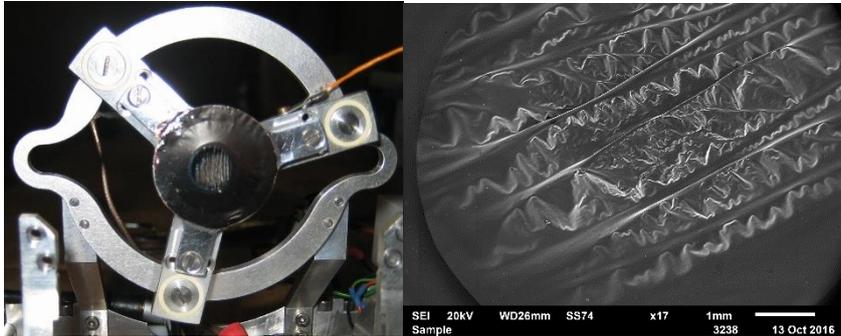

*Fig. 2. upper : Titanium target mounted on the plunger set-up after irradiation; lower: scanning electron microscope (electron 20kV) photograph of the target with magnification *17*

Several causes were then questioned in order to explain the observations. Indeed, the damage can originate from:
- Mechanical stress induced by the fabrication process or the method of stretching and fixation on the frame.
- Chemical reaction with air in the time between fabrication and irradiation under vacuum, some material can oxidize easily and become fragile.
- Interaction with another material in the case of a backing.
- Irradiation inducing heating, phase change or damage of the material.

The last three causes can modify the properties of the material to unexpected values of some parameters (elasticity, conductivity, melting temperature, …) which may make the target delicate to sustain the beams.

In plunger systems, the targets undergo exceptional mechanical strain because they have to be stretched and fixed over a small solid cone. This requirement was verified and qualified before irradiation. But as observed on Fig. 2, the rolling method used to manufacture the target might have induced during irradiation a preferential direction of the wrinkle-like structures of the Ti targets.

In all cases studied, the target temperature at thermodynamically equilibrium were calculated by taking into account the properties of the beam (energy, intensity, spot size) and the target material (emissivity, thermal conductance, thickness). This temperature was estimated considering cooling by radiation and conduction. The values obtained were lower than the fusion temperature of the material reported in literature.

Then the damage formation resulting from a nanometric, dense and shortly created electronic excitation in the material, was investigated using the Inelastic Thermal Spike model (i-TS) [3-6].

Indeed, the passage of a fast ion through a foil creates a "heat spike" along its path, in which the temperature may be briefly extremely high, even if the average temperature of the foil remains low. Within this model, this paper reports on the temperature distribution in time along the beam track calculated for the different projectile/target systems under consideration. The results of the model are compared to the temperature needed for the material to reach the melt phase when the track formation appears in the material.

## 2 Thermal Spike Model

### 2.1 Description

The i-TS model is one of the most powerful tool to describe damage induced by swift heavy ions and is most widely applied to any target material either metallic, semi-conductor or insulators.

The model reproduces the electronic excitation of the material through a four step process: first, the incident ions transfer their energy to the electrons of the target by ion-electron collisions. Second, electron-electron collisions share this energy with other cold electrons. Third, the energy is transferred to the lattice by electron-phonon coupling. Forth, the energy dissipates among the atoms. This process induces a spike along the ion trajectory. Regarding the timeline, illustrated on figure 2.1 from [7], the energy is deposited in the electrons within $10^{-16}$ to $10^{-15}$ s and then transferred to the lattice atoms within $10^{-13}$ to $10^{-11}$ s.

Mathematically this model is based on specific solutions of two coupled differential equations (1) and (2) governing the heat diffusion to the electrons and atomic subsystems versus time $t$ and space $r$ (in cylindrical geometry) and their exchange via the electron-phonon coupling:

$$C_e(T_e)\frac{\partial T_e}{\partial t} = \frac{1}{r}\frac{\partial\left[rK_e(T_e)\frac{\partial T_e}{\partial r}\right]}{\partial r} - g(T_e - T_a) + A(r) \qquad (1)$$

$$C_a(T_a)\frac{\partial T_a}{\partial t} = \frac{1}{r}\frac{\partial\left[rK_a(T_a)\frac{\partial T_a}{\partial r}\right]}{\partial r} + g(T_e - T_a) \qquad (2)$$

In these equations, the parameters $T_{e,a}(r,t)$, $C_{e,a}(r,t)$ and $K_{e,a}(r,t)$ are the temperature, specific heat and thermal conductivity of the electronic (*e*) and atomic (*a*) subsystem, respectively. $A(r)$ is the energy deposited into the electronic subsystem supplied by the incident ion by ballistic collisions at radius *r*. The integration of $A(r)$ over space gives the total stopping power $dE/dx$. The only free parameter in this model is the electron-phonon coupling strength *g*.

Due to the very short time of energy deposition ($10^{-13} – 10^{-12}$ s), the model calculations are performed within a superheating scenario [8], i.e. that increase of temperature does not stop when reaching the melting or the boiling temperature.

The track radii are associated with the cylinder in which the energy deposited on the atoms surpasses a specific energy. This limit, $E_{SH}$, is defined as the energy to reach the melting temperature, $E_{fus}$, plus the energy required to make the solid-liquid phase change (latent heat of fusion, $E_{lat}$). This criterium is used since the superheating scenario of the target is the result of a very rapid heating stage ($\approx 10^{-13}$ s) [7]. As heat capacity of a material is a function of its internal energy and temperature (equation (3)), the $T_{SH}$ temperature is deduced from the integration of the specific heat from $T_0=0°K$ to $T_{SH}$ which equals the $E_{SH}$ energy.

$$C_a = \frac{dE}{dT} \qquad (3)$$

## 2.2 Application to experiments

### 2.2.1 Experimental conditions

The experiments were performed at GANIL using heavy ion beams around Coulomb barrier energy impinging on targets as listed in Table 1.

Cases 1, 2 and 3 correspond to experiments using the plunger device mounted in the VAMOS target chamber at an angle of 45° with respect the beam axis. A magnesium degrader slows down the recoiling nuclei produced in the first target placed at an adjustable distance downstream. The first targets were self-supporting and manufactured by rolling. In case 2, a layer of copper was evaporated on the titanium in order to increase its thermal conductivity. The copper layer was facing the beam.

These systems are compared to cases 4 and 5 using either a similar target (Ni) in direct kinematics with a "light" ion beam ($^{50}$Cr) or a target of $^{54}$Fe irradiated with a heavy ion $^{124}$Xe beam at comparable velocities.

In case 4, a gold foil was used to slow down the beam before reacting in the nickel target.

In case 5, iron was evaporated on a thin carbon layer, four targets were irradiated and revealed various behaviour.

The effective material thicknesses are given taking into account the angle of the target with respect to the incoming beam (45° in cases 1, 2 and 3 and 0° for cases 4 and 5).

*Table 1. Experimental properties of the beam (energy and intensity) and targets (effective thickness).*

| |
|---|
| **Experiment 1:** $^{238}$U$^{32+}$(6.2 MeV/u; 1 pnA) + $^{64}$Ni (2.6 mg/cm²) + $^{nat}$Mg (5.2 mg/cm²) <br> No problem of target, Mg degrader slightly deformed |
| **Experiment 2:** $^{238}$U$^{32+}$ (6.76 MeV/u; 0.1 pnA) + Cu-$^{50}$Ti (0.6 - 2.1 mg/cm²) + $^{nat}$Mg (4.5 mg/cm²) <br> Target seriously deformed quickly $I_{max}$= 3 nAe = 0.1 pnA |
| **Experiment 3:** $^{238}$U$^{32+}$ (6.76 MeV/u; 0.07pnA) + $^{50}$Ti (2.1 mg/cm²) + $^{nat}$Mg (4.5 mg/cm²) <br> Target seriously deformed quickly $I_{max}$= 2 nAe = 0.07 pnA |
| **Experiment 4:** $^{50}$Cr$^{8+}$(4 MeV/u; 10 pnA) + Au (2.4 mg/cm²) + $^{58}$Ni (7-10 mg/cm²) <br> No problem at 30 nAe |
| **Experiment 5:** $^{124}$Xe$^{39+}$(4 MeV/u; 1 pnA) + C -$^{54}$Fe (0.03 - 0.2 mg/cm²) <br> 3 targets have broken below 6 nAe and 1 sustained up to 80 nAe |

*2.2.2 Inputs of the i-TS calculations*

The initial temperature at equilibrium ($T_{eq}$ in Table 2) of the target is determined by the beam intensity, energy and spot size. It was calculated considering cooling by radiation and conduction through the target holder. The external surfaces radiate towards a black body at room temperature and if necessary, the mutual radiation between two closed targets were considered. The range of temperature is estimated with a beam spot represented by a Gaussian distribution with σ = 1mm and emissivity values ranging from 0.1 to 0.9 [9-10].

The measured thermal conductivity ($K_s$) of a metal corresponds to the electron conductivity taking into account all the properties of the electron gas. Consequently, using $K_s$ as the input of the metal thermal conductivity, the value of the g factor should be near to the one calculated with a number of electrons equal to the number of atoms [5].

The value of g was determined by fitting the track size confirming the validity of this hypothesis, apart for Ti for which the g factor is twice larger than the expected one. Then the electron-phonon coupling factors g are from [5] and were confirmed or adjusted from detailed study of experimental sputtering yields in nickel [6], titanium [11], and iron [11] materials.

*Table 2. Parameters of the materials for the i-TS simulations*

| Material experiment | dP [*mwatts*] | $T_{eq}$ [°K] | DE/dx [*keV/Å*] |
|---|---|---|---|
| **Mg** | | | |
| 1 | 540 | 850±100 | 1.9 |
| 2/3 | 48 | 500±100 | 1.9 |
| **Ni** | | | |
| 1 | 200 | 850±100 | 7.8 |
| 4 | 1700 | 1150±100 | 1.5 |
| **Cu** | | | |
| 2 | 4 | 500±100 | 7.2 |
| **Ti** | | | |
| 2/3 | 18 | 450±100 | 4 |
| **Au** | | | |
| 4 | 250 | 750±100 | 2 |
| **Fe** | | | |
| 5 | 10 | 450±50 | 4 |

| Material | $g*10^{10}$ [W/cm$^3$/K] | $n*10^{22}$ [e$^-$/cm$^3$] |
|---|---|---|
| **Mg** | 16.8 [5] | 4.3 [5] |
| **Ni** | 107 [5-6] | 9.14 [5-6] |
| **Cu** | 12.7 [5] | 8.5 [5] |
| **Ti** | 1000 [11] | 5.7 [5] |
| **Au** | 2.3 [5] | 5.9 [5] |
| **Fe** | 119 [11, 5] | 8.9 [5] |

The thermodynamically temperature at equilibrium was well below the fusion temperature of the material for all experiments except for the first one (Mg) where it could have been close.

*2.2.3 Results*

For comparison, the simulations with i-TS model were done at 300 °K and some temperatures around $T_{eq}$ reported in Table 2 in all cases.

The results of the calculations are given in Fig. 3 to Fig. 8 where the lattice temperatures are plotted versus time along ion path. The temperatures are represented by full, dashed and dashed-dotted lines for the initial temperature of the material of 300°K, mean $T_{eq}$ (written in bold and underlined) and $T_{eq} \pm 100$ respectively.

The limit temperature ($T_{SH}$) corresponds to the energy necessary to melt the material defined as the melting temperature plus the corresponding latent heat of fusion, their values are reported in Table 3 and plotted in the figures where they could be reached together with the corresponding fusion temperature.

Table 3. "Super-Heating" and fusion temperatures [°K] of materials

|  | Mg | Ni | Cu | Ti | Au | Fe |
|---|---|---|---|---|---|---|
| $T_{SH}$ | 1260 | 2250 | 1900 | 2580 | 1790 | 2180 |
| $T_{fus}$ | 923 | 1728 | 1358 | 1941 | 1337 | 1811 |

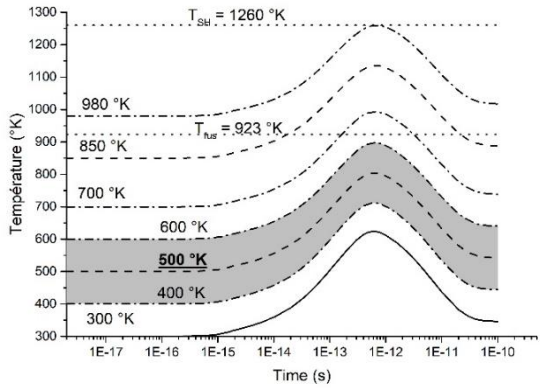

Fig. 3. Evolution of the lattice temperature of the Mg degrader as a function of time along the ion path for cases 1, 2 and 3 (filled in grey).

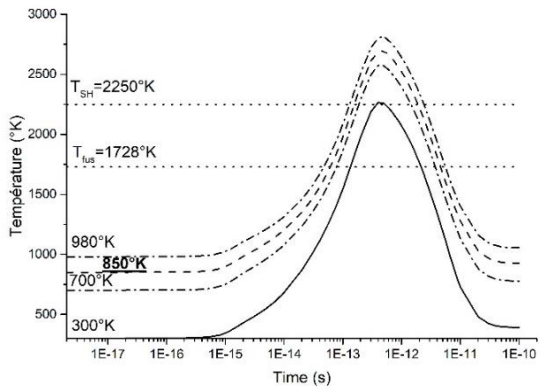 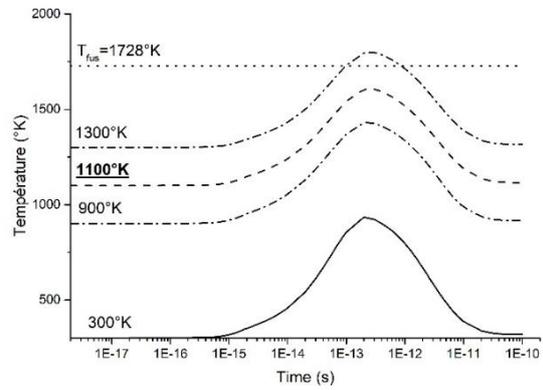

Fig. 4. Evolution of the lattice temperature of the Ni target as a function of time along the ion path for cases 1 (upper) and 4 (lower).

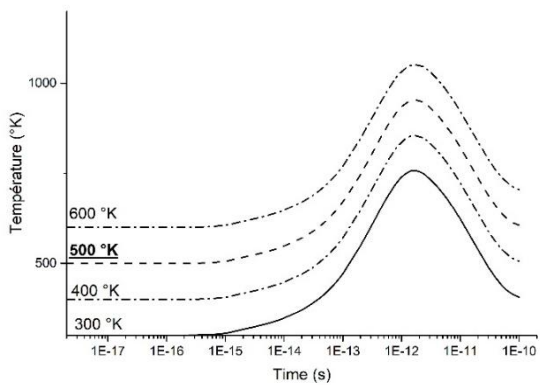

Fig. 5. Evolution of the lattice temperature of the Cu target as a function of time along the ion path for cases 2.

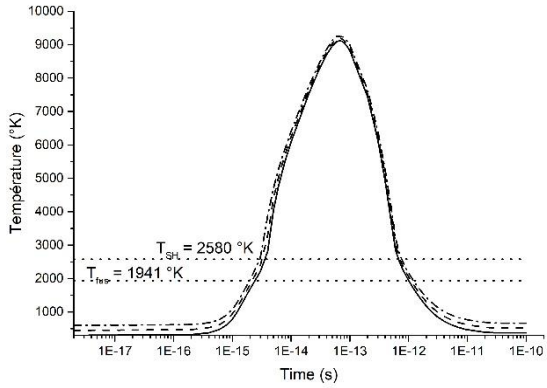

*Fig. 6. Evolution of the lattice temperature of the Ti target as a function of time along the ion path for cases 2 and 3.*

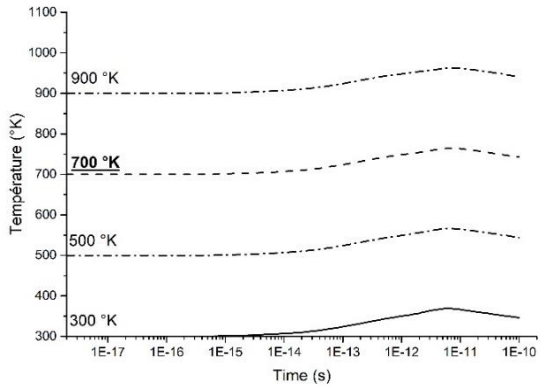

*Fig. 7. Evolution of the lattice temperature of the Au target as a function of time along the ion path for cases 4.*

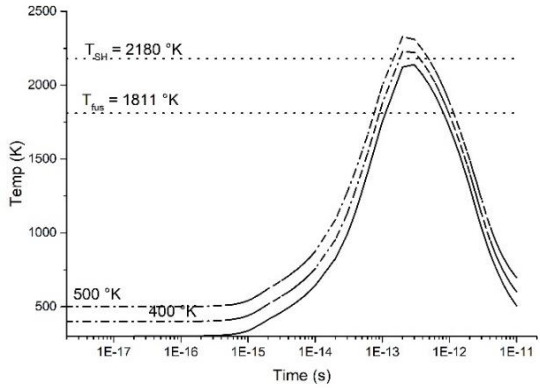

*Fig. 8. Evolution of the lattice temperature of the Fe target as a function of time along the ion path for cases 5*

## 3 Discussion

The temperature distributions over time for the different systems show various features.

First the temperature is maximum at a different time according to the material, in the ranges [$10^{-14}$-$10^{-13}$ s], [$10^{-13}$-$10^{-12}$ s] or [$10^{-12}$-$10^{-11}$ s] for Ti, "Fe, Au, Mg, Ni" or Cu respectively. This tendency is directly linked to the electron-phonon strength: the larger it is, the smaller are the heating and cooling times.

Secondly, the temperature rise varies according to either the material or the initial conditions. It ranges from ≈65°K for Au (Fig. 7) (or ≈300°K for Mg (Fig. 3), ≈450°K for Cu (Fig. 5)) to ≈8700°K for Ti (Fig. 6) and for Ni targets from ≈560 °K in experiment 4 (Fig. *4* lower) to ≈1880 °K in experiment 1 (Fig. *4* upper).

In experiment 4, the nickel and gold targets did not show any damage, and the model shows that the temperature increase due to spike did not reach the limit temperature ($T_{SH}$).

Compared to the first experiment, where the electronic stopping power of the Ni target was higher (7.8 keV/Å), its temperature increase could have influenced the Mg degrader temperature at equilibrium above the fusion temperature, explaining that it was slightly deformed as observed.

In the second experiment, the temperature distribution of the copper material (Fig. 5) is well below its limit temperature, by contrast with the titanium response to the spike where $T_{SH}$ is reached fast, what could explain the rapidly degradation of the targets (Ti alone or Ti with Cu evaporated) at very low beam intensity.

As explained in [12] and regarding Fig. 9 for Ti targets, the latent track radius, defined as the radius of a cylinder of matter in which the energy necessary to melt is exceeded, is calculated to be about 8 nm. This value is in accordance with the one measured in [13] and calculated in [11] at dE/dx = 40 keV/nm. The radius of 8 nm corresponds to a fluence of $1.2*10^{12}$ particle per cm² inducing a first modification of the material, in other words before overlapping of spikes. This fluence is reached within 1 to 16 minutes with a beam of 0.1 pnA and a spot radius of 1 to 4 mm. Track overlapping can induce phase change in the titanium [14] which can affect drastically the target stability.

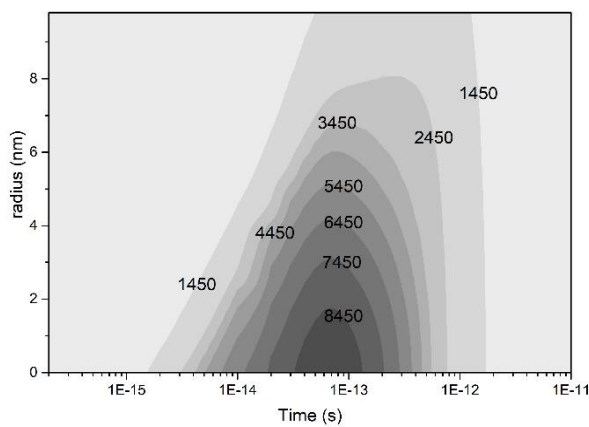

*Fig. 9. Temperature distribution of the Ti target as calculated with the Thermal Spike Model.*

The temperature rise is pronounced for Fe (≈1800 °K, Fig. 8), and the higher value is near $T_{SH}$ independent of the initial temperatures. However, the four iron targets used did not behave similarly under irradiation, one of them could sustain the beam up to an intensity of 80 nAe, the other ones broke at low intensity and within few hours. Apart from the potential "thermal spike" effect, one can speculate that other contributions to the damage of targets originates from the carbon/iron interface (carbon used as a backing) because of their different thermal expansion coefficients and/or the high sensitivity of iron to oxidation when it was on air before irradiation.

## 4 Conclusion

The thermal spike model, applied to some projectile-target combinations allows to understand that some metallic targets were deformed due to a fast and local temperature increase while others were insensitive. The model enables to predict reliably the sensitivity of targets to the electronic slowing down of heavy ion by considering fundamental parameters such as the energy loss and the electron-phonon coupling factor determined by some sputtering yields experimental studies. Preliminary estimations with this model prior to experiments will help to adjust some parameters, i.e. the beam energy and target material to prevent the damages.